\documentclass[runningheads]{llncs}
\usepackage[T1]{fontenc}
\usepackage{graphicx,multirow,amssymb}
\usepackage{marvosym}
\usepackage{color,xcolor}
\usepackage[colorlinks,
linkcolor=blue,
anchorcolor=blue,
urlcolor=blue,
citecolor=blue]{hyperref}
\begin{document}
\title{LesionPaste: One-Shot Anomaly Detection for Medical Images}
%\author{Anonymous}
\author{Weikai Huang\inst{1} \and
Yijin Huang\inst{1,2} \and
Xiaoying Tang$^{1(\textrm{\Letter)}}$}
%\textsuperscript{\Letter}

\authorrunning{W. Huang et al.}
%\authorrunning{Anonymous}
%\institute{Anonymous Organization\\
%\email{**@******.***}}
\institute{Department of Electronic and Electrical Engineering, \\
	Southern University of Science and Technology, Shenzhen, China\\
\email{tangxy@sustech.edu.cn}\and
School of Biomedical Engineering, \\
University of British Columbia, Vancouver, British Columbia, Canada}

\maketitle              % typeset the header of the contribution
\begin{abstract}
Due to the high cost of manually annotating medical images, especially for large-scale datasets, anomaly detection has been explored through training models with only normal data. Lacking prior knowledge of true anomalies is the main reason for the limited application of previous anomaly detection methods, especially in the medical image analysis realm. In this work, we propose a one-shot anomaly detection framework, namely LesionPaste, that utilizes true anomalies from a single annotated sample and synthesizes artificial anomalous samples for anomaly detection. First, a lesion bank is constructed by applying augmentation to randomly selected lesion patches. Then, MixUp is adopted to paste patches from the lesion bank at random positions in normal images to synthesize anomalous samples for training. Finally, a classification network is trained using the synthetic abnormal samples and the true normal data. Extensive experiments are conducted on two publicly-available medical image datasets with different types of abnormalities. On both datasets, our proposed LesionPaste largely outperforms several state-of-the-art unsupervised and semi-supervised anomaly detection methods, and is on a par with the fully-supervised counterpart. To note, LesionPaste is even better than the fully-supervised method in detecting early-stage diabetic retinopathy.

\end{abstract}                   
\section{Introduction}
In recent years, deep learning has achieved great success in the field of medical image analysis \cite{survey}. However, the effectiveness of deep representation learning techniques, such as convolutional neural networks (CNNs), is severely limited by the availability of the training data. Most of the disease detection methods in the medical image field are fully supervised and heavily rely on large-scale annotated datasets \cite{review1}. In general, acquiring and manually labeling abnormal data are more challenging and more expensive than normal data. Thus, a vast majority of anomaly detection methods in computer vision have been focusing on unsupervised learning through training detection models using only normal data, assuming no access to abnormal data at the training phase \cite{Student-teacher,knowledge_distil,inpainting}.
%Anomaly detection has been proposed to detect the abnormal samples input during the test by identifying the data points that deviate significantly from the majority of data samples' behavior \cite{review2}.
%\begin{figure*}[htb]
%	\centering
%	\includegraphics[width=8cm]{fig1.pdf}
%	%  \vspace{2.0cm}
%	\caption{Three different anomalous medical images and their annotated lesions}
%	\label{figure1}
%\end{figure*}

Typically, these methods use normal samples to train models to learn normality patterns and declare anomalies when the models have poor representation of specific test samples \cite{review2}. For instance, training an autoencoder that reconstructs normal samples by minimizing the reconstruction error can be used to detect anomalies when the reconstruction error is high in testing \cite{dream,p-net}. Generative models aim to learn a latent feature space that captures normal patterns well and then defines the residual between real and generated instances as the anomaly score to detect anomalies \cite{gan2,f-AnoGAN}. Recently, some approaches have been explored by synthesizing anomalous samples. For example, CutPaste performs data augmentation by cutting image patches and pasting them at random locations in normal images to serve as coarse approximations of real anomalies for anomaly detection \cite{Cutpaste}. DREAM synthetically generates anomalous samples to serve as the input to its reconstruction network and calculates anomaly scores based on the reconstruction results \cite{dream}. Since there is no prior knowledge of the true anomalies, these methods generally use very simple and rough methods to synthesize anomalous samples, although of high effectiveness. A major limitation is that the anomaly score defined as the pixel-wise reconstruction error or the generative residual relies heavily on the assumption on the anomaly distribution \cite{NIPS2019}. Therefore, these methods may not be sufficiently robust and generalizable in discriminating anomalies in real-life clinical practice. 

In such context, we propose LesionPaste, a one-shot anomaly detection (OSAD) method, which is the extreme case of few-shot anomaly detection \cite{few-short1,few-short2}. Namely, we train an anomaly detection network with only one annotated anomalous sample. Requiring only a single labeled anomalous sample, LesionPaste is highly flexible and accommodates well various settings even for rare diseases or other unusual instances. Our goal is to make use of the prior knowledge of true anomalies to synthesize artificial anomalous samples, at the cost of annotating anomalies in only a single anomalous sample.

In LesionPaste, we first choose one annotated anomalous image and extract all lesion patches. Then, data augmentation is applied to the extracted lesion patches to construct a lesion bank. Afterwards, MixUp is employed to paste lesion patches from the lesion bank to normal images to synthesize artificial anomalous images. Finally, we train an anomaly detection network by discriminating synthesized anomalous images from normal ones. The performance of our proposed LesionPaste is evaluated on two publicly-available medical image datasets with different types of abnormalities, namely EyeQ \cite{eyeq} and MosMed \cite{MosMedData}.

The main contributions of this work are two-fold: (1) We propose a novel OSAD framework, namely LesionPaste, to utilize the prior knowledge of true anomalies from a single sample to synthesize artificial anomalous samples. To the best of our knowledge, this is the first work that synthesizes artificial anomalous samples using real anomalies from a single sample, for a purpose of anomaly detection. (2) We comprehensively evaluate our framework on two large-scale publicly-available medical image datasets including fundus images and lung CT images. The superiority of LesionPaste is established from the experimental results. The source code is available at \url{https://***/***}.
\section{Methods}
%\subsection{Synthesis Strategy For DR datasets}
%There are four different lesions on DR fundus images: microaneurysms (MA), hemorrhages (HE), soft exudates (SE) and hard exudates (EX), of which MA is the smallest and hardest to classify \cite{DR_Lesion}. According to the clinical definition of DR grading, DR grade 1 only includes MA, DR grade 2 includes MA and HE, and the more severe DR contains more types of lesions. To let CNN learn more hard classify samples to improve performance, we design a strategy for artificial DR image synthesis: 80\% of normal fundus images synthesize with MA only, 10\% synthesize with MA and HE, 5\% synthesize with MA, HE and SE, and the remaining 5\% synthesize with all types of lesions.
\subsection{Construction of Lesion Bank}
\begin{figure*}[htbp]
	\centering
	\includegraphics[width=12cm]{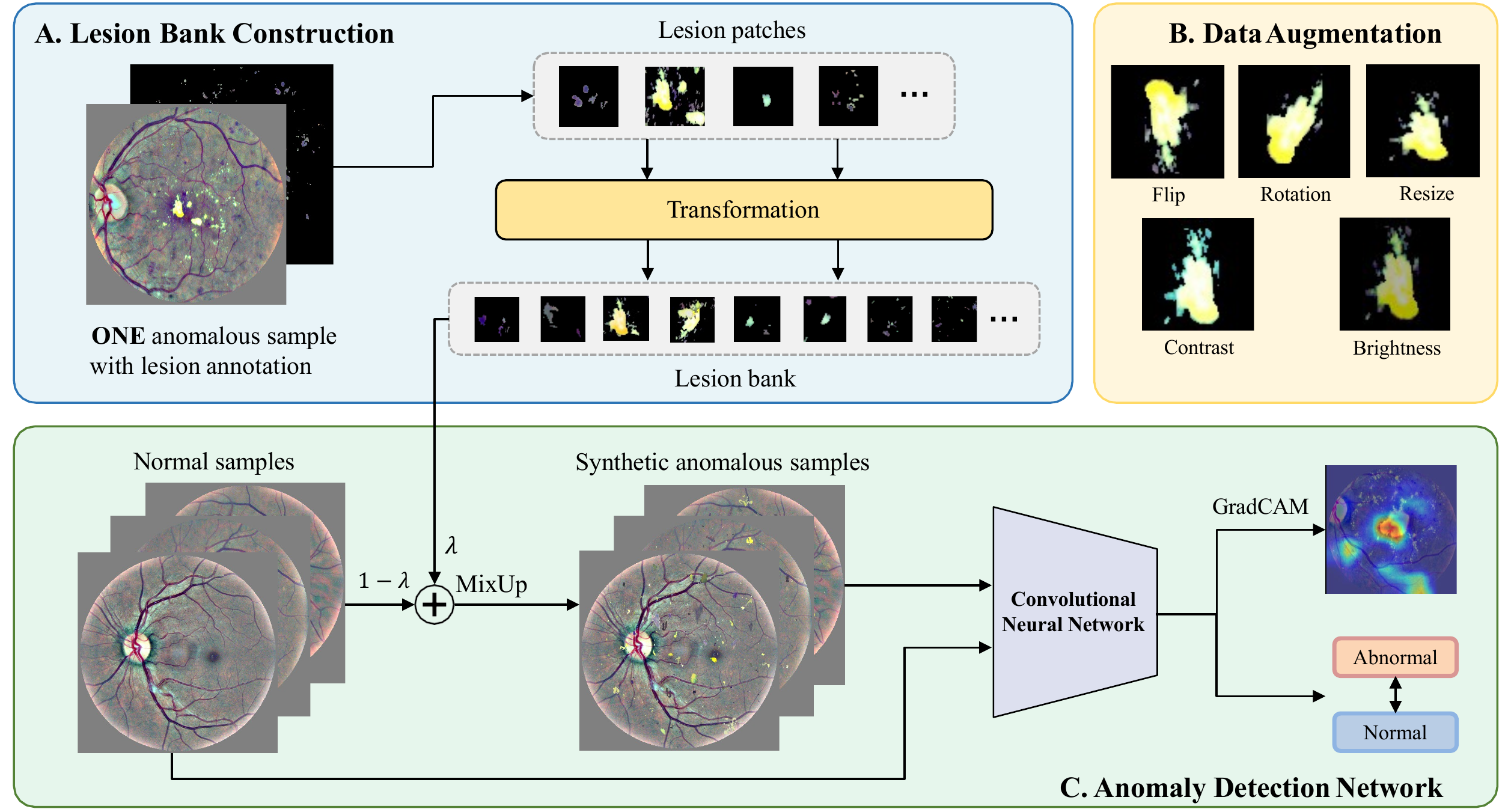}
	%  \vspace{2.0cm}
	\caption{The proposed LesionPaste framework. In part A, lesion patches are extracted from an annotated anomalous sample. Then, the lesion patches are processed by a transformation block involving data augmentation operations in part B to construct a corresponding lesion bank. In part C, the transformed lesion patches in the lesion bank are randomly pasted to the normal samples via MixUp, yielding synthetic anomalous samples. Then a CNN is trained to detect the anomalies.}
	\label{pipeline}
\end{figure*}

\begin{figure*}[htbp]
	\centering
	\includegraphics[width=6.4cm]{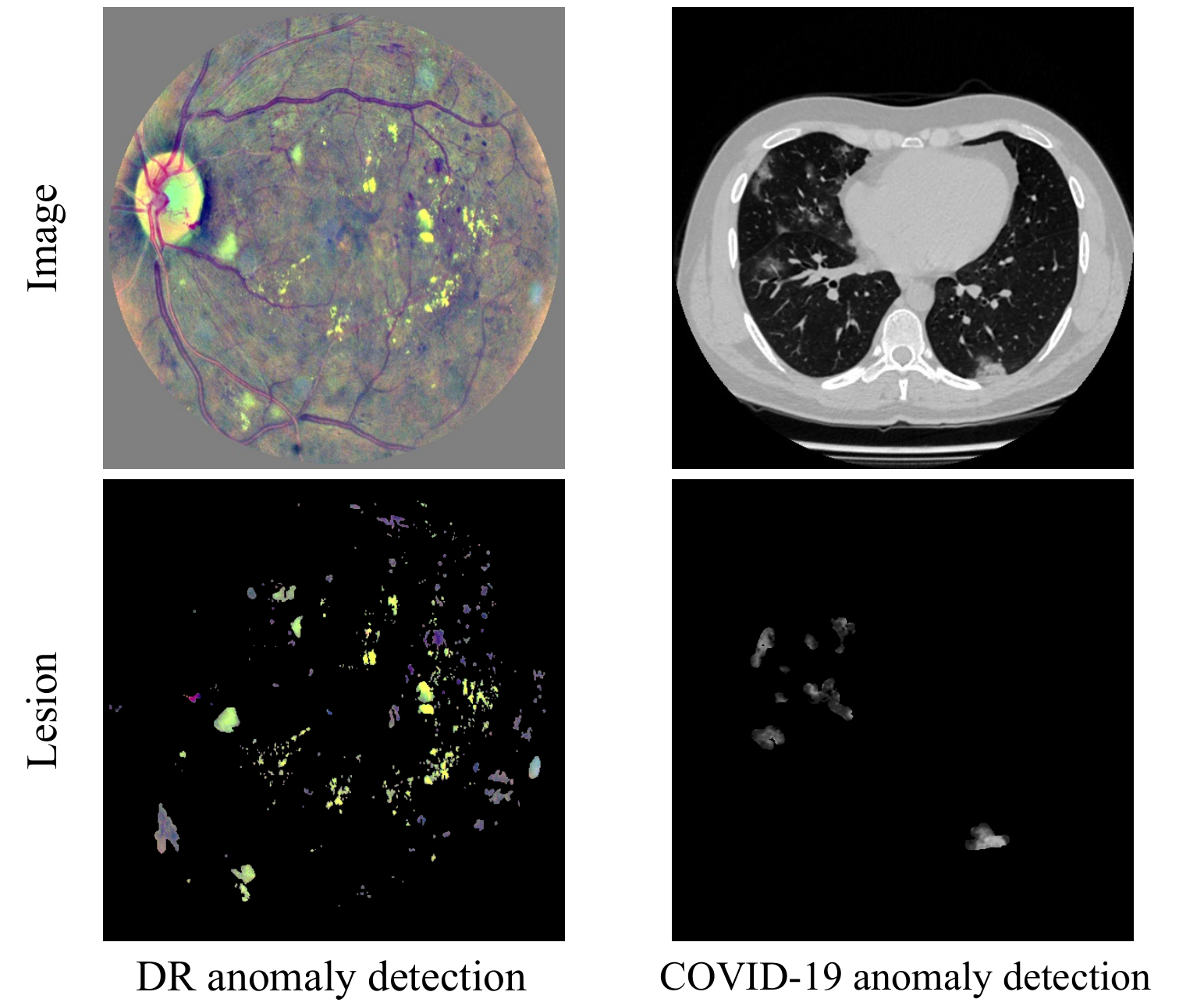}
	%  \vspace{2.0cm}
	\caption{The single annotated anomalous data for each of our two tasks.}
	\label{image_lesions}
\end{figure*}

As depicted in Fig.~\ref{pipeline}A, we first choose one annotated anomalous image and extract all lesions based on the pixel-wise lesion annotation. For the two anomaly detection tasks investigated in this work, the single annotated anomalous data are illustrated in Fig.~\ref{image_lesions}. After lesion extraction, the Connected Component Labeling algorithm \cite{CCL} is adopted to identify each isolated lesion region from which a corresponding lesion patch is extracted.

Following that, random resampling with repetition is carried out to select the lesion patches to be pasted. The number $N$ of the to-be-pasted lesion patches is also randomly generated with $N \sim U(1, 1.5N_L)$, where $N_L$ is the total number of the isolated lesion regions. The selected lesion patches are then sent to a subsequent transformation block for lesion patch augmentation to construct a lesion bank, for a purpose of synthesizing more diverse anomalies. Data augmentations are conducted as shown in Fig.~\ref{pipeline}B, including flipping, rotation, resizing, contrast, and brightness changing, to generalize our anomalies from a single sample to various unseen anomalies during testing.
\subsection{Synthesis of Anomalous Samples}
We randomly sample a number of lesion patches from the lesion bank and paste them at random positions in the normal images to synthesize artificial anomalous images. Each normal image is used to generate one corresponding artificial anomalous image.

The MixUp technique is initially proposed as a simple data augmentation method in \cite{mixup} to regularize model complexity and decrease over-fitting in deep neural networks by randomly combining training samples. Extensive experiments have shown that MixUp can lead to better generalization and improved model calibration. As such, to have the pasted lesion fuse more naturally with the normal image, random MixUp is employed when we paste a lesion patch $L$ to a normal image $I$. The image after MixUp $I_{MU}$ is defined as
\begin{equation}
I_{MU}=(1-\lambda) \left(M \odot I\right)+\lambda\left(M \odot L\right)+\bar{M} \odot I,
\end{equation}
where $M$ is the binary mask of the lesion patch, $\bar{M}$ is the inverse of $M$, $\odot$ denotes the pixel-wise multiplication operation and $\lambda \sim U(0.5, 0.8)$.
\subsection{Anomaly Detection Network}
After generating the artificial anomalous samples, as shown in Fig.~\ref{pipeline}C, together with the normal data, a CNN can be trained to detect anomalies. VGG16 initialized with ImageNet parameters is adopted as our backbone model.

Let $\mathcal{N}$ denote a set of normal images, $\mathbb{C} \mathbb{E}(\cdot, \cdot)$ a cross-entropy loss, $f(\cdot)$ a binary classifier parameterized by VGG16 and AP$(\cdot)$ a LesionPaste involved augmentation operation, the training loss function of the proposed LesionPaste framework is defined as
\begin{equation}
\mathcal{L}_{\mathrm{AP}}=\mathbb{E}_{I \in \mathcal{N}}\{\mathbb{C} \mathbb{E}(f(I), 0)+\mathbb{C} \mathbb{E}(f(\mathrm{AP}(I)), 1)\}. 
\end{equation}

\subsection{Implementation Details}
In both training and testing phases, images are resized to be 256$\times$256 and the batch size is set to be 64. We adopt an SGD optimizer with a momentum factor of 0.9, an initial learning rate of 0.001, and the cosine decay strategy to train the network. All network trainings are performed for 50 epochs on EyeQ and 100 epochs on MosMed, with fixed random seeds. All compared models and the proposed LesionPaste framework are implemented with Pytorch using NVIDIA TITAN RTX GPUs.
\section{Experiments and Results}
\subsection{Datasets}
%\subsubsection{EyePACS.} EyePACS \cite{EyePACS} provides 35k/11k/43k fundus images for training/ validation/testing. According to the severity of diabetic retinopathy (DR), images are classified into five grades: 0 (normal), 1 (mild), 2 (moderate), 3 (severe), and 4 (proliferative). Images of grades 1-4 are considered anomalies. All normal images in the training set are used to train our OSAD model and all images in the testing set are used for evaluation.

\subsubsection{EyeQ.} EyeQ \cite{eyeq} is a subset of the famous EyePACS \cite{EyePACS} dataset focusing on diabetic retinopathy (DR), consisting of 28792 fundus images with quality grading annotations. The quality of each image is labeled as "good", "usable", or "reject". In our experiments, we remove images labeled as either "usable" or "reject", ending up with 7482/865/8471 fundus images for training/validation/testing. According to the severity of DR, images in EyeQ are classified into five grades: 0 (normal), 1 (mild), 2 (moderate), 3 (severe), and 4 (proliferative) \cite{DR_dataset}. The class distribution of the training data is shown in Fig. \url{A1} of the appendix. Images of grades 1-4 are all considered as abnormal. All normal images in the training set are used to train LesionPaste and all images in the testing set are used for evaluation. All fundus images are preprocessed \cite{preprocess} to reduce heterogeneity as much as possible, as shown in Fig. \url{A1} of the appendix. 

\subsubsection{IDRiD.} IDRiD \cite{idrid} consists of 81 DR fundus images, with pixel-wise lesion annotations of microaneurysms (MA), hemorrhages (HE), soft exudates (SE) and hard exudates (EX) \cite{DR_Lesion} (see Fig. \url{A2} of the appendix). In our LesionPaste, the lesions of a single fundus image from IDRiD are used as the true anomalies for DR anomaly detection. According to the clinical definition of DR grading, DR of grade 1 only contains MA, DR of grade 2 contains MA and HE, and DR of more severe grades contains more types of lesions. To enhance the performance of the detection CNN, we design a specific strategy for synthesizing the anomalous DR images: 80\% of the normal fundus images are Mixed-Up with lesion patches containing MA only, 10\% with lesion patches containing both MA and HE, 5\% with lesion patches containing MA, HE and SE, and the remaining 5\% with lesion patches containing all the four types of lesions (MA, HE, SE, and EX). 
%Diabetic retinopathy (DR), one of the microvascular complications of diabetes, can lead to visual impairment and irreversible blindness in patients \cite{DR_dataset}. Color fundus images are the most commonly used modality for DR identification clinically, and it can clearly reveal the presence of different lesions associated with DR grading \cite{DR_Lesion}.
\begin{figure*}[htbp]
	\centering
	\includegraphics[width=12cm]{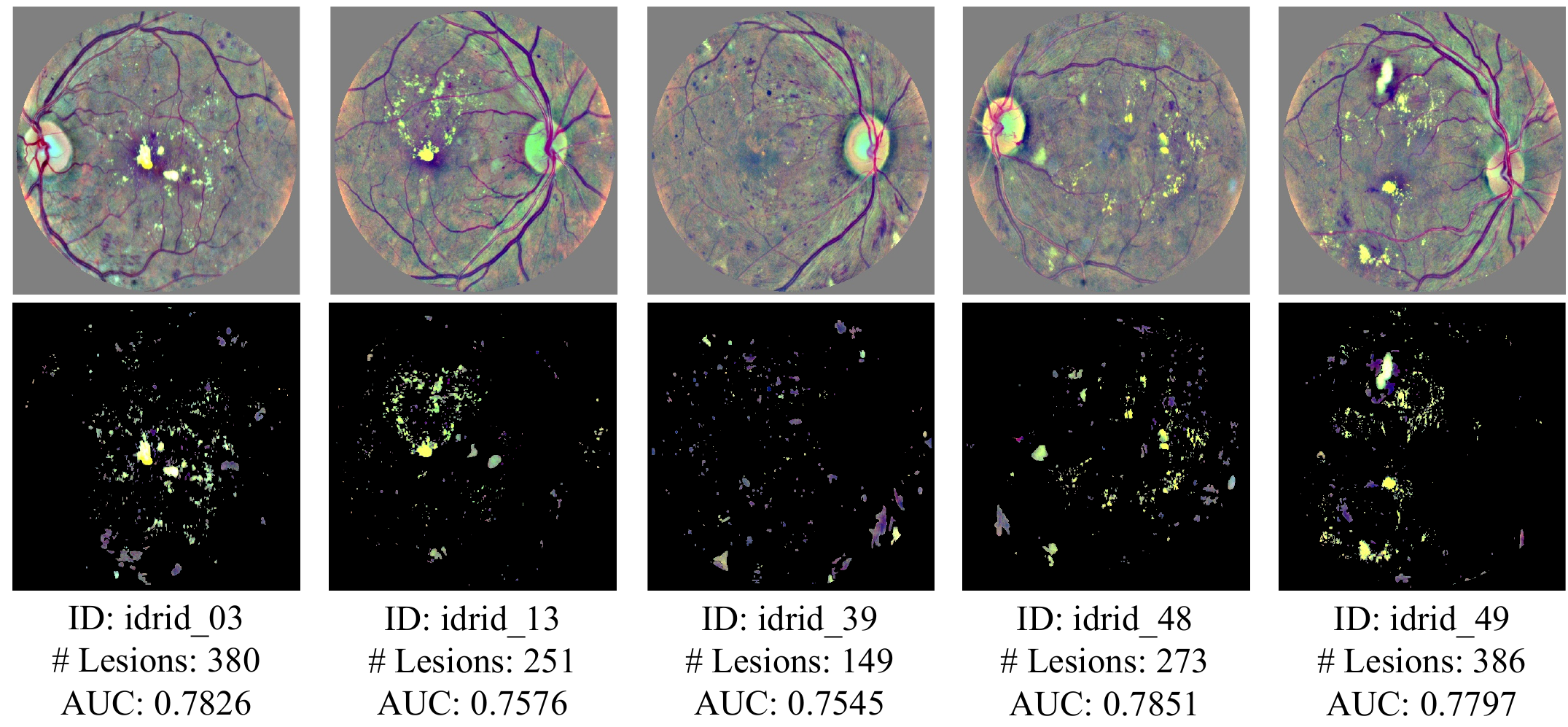}
	%  \vspace{2.0cm}
	\caption{The impact of using different annotated samples. ID denotes the image index in IDRiD , $\#$Lesions denotes the number of isolated lesion regions, and AUC is reported for detecting all DR images combining grades 1-4.}
	\label{IDRiD_sample}
\end{figure*}
\subsubsection{MosMed.} MosMed \cite{MosMedData} contains human lung CT scans with COVID-19 related findings, as well as some health samples. A small subset have been annotated with pixel-wise COVID-19 lesions. CT slices containing COVID-19 lesions are considered as abnormal, ending up with 759 slices. A total of 2024 normal CT slices are selected and extracted from 254 health samples. For the anomaly detection task on this dataset, 5-fold cross-validation is used for evaluation. All CT images are preprocessed by windowing with a window level of -300 HU and a window width of 1400 HU to focus more on lung tissues \cite{CT_window}.

%\subsubsection{Kvasir.} Kvasir dataset \cite{Kvasir} contains 10k endoscopic images and 10 different classes, with 1,000 image examples per class. Images of polyps in Kvasir, which have corresponding segmentation ground truth \cite{Kvasir_seg}, are considered anomalies. 5-fold cross-validation is used for evaluation.

\subsection{Evaluation Metric} The anomaly detection performance is evaluated using a commonly-employed metric, namely the area under the curve (AUC) of receiver operating characteristic (ROC), keeping consistent with previous anomaly detection works \cite{Cutpaste,few-short1,dream}.

\subsection{Ablation Studies on EyeQ}

\subsubsection{Different Annotated Samples.}
In this experiment, we only use the original lesion patches with no data augmentation, and evaluate the performance of LesionPaste with different annotated fundus images of IDRiD. In Fig.~\ref{IDRiD_sample} and Table \url{A1} of the appendix, we show the results of randomly selecting five different images from IDRiD for lesion extraction. Apparently, the difference in the single annotated anomalous image does not affect the anomaly detection performance of our LesionPaste, identifying the robustness of our proposed pipeline. We choose a representative image, idrid\_48, as the single annotated fundus image for all subsequent experiments on EyeQ.

\begin{table}[htbp]
	\caption{The impact of different data augmentation operations on EyeQ. DR images of grades 1-4 are all considered as abnormal. Color distortion only adjusts the hue and saturation and does not change brightness.}
	\centering
	\begin{tabular*}{\hsize}{@{}@{\extracolsep{\fill}}ccccccc}
		\hline Color distortion  & Flip& Contrast & Rotation&Resize   & Brightness& AUC\\
		\hline$\checkmark$ & & & &&& $0.7806$\\
		& $\checkmark$ & & &&& $0.7803$  \\
		& & $\checkmark$ & &&&  $0.7841$\\
		& & & $\checkmark$ & &&$0.7864$ \\
		&& & & $\checkmark$  &&$0.7896$\\
		&& & & & $\checkmark$  &$0.7924$ \\
		\hline&& & &  $\checkmark$& $\checkmark$  &$0.7965$\\
		&& &  $\checkmark$& $\checkmark$ & $\checkmark$  &$0.8019$ \\
		&& $\checkmark$ &  $\checkmark$& $\checkmark$ & $\checkmark$  &$0.8105$ \\
		& $\checkmark$&  $\checkmark$& $\checkmark$ &  $\checkmark$& $\checkmark$  &$\textbf{0.8126}$\\
		$\checkmark$& $\checkmark$&  $\checkmark$& $\checkmark$ &  $\checkmark$& $\checkmark$  &$0.8053$\\
		\hline
	\end{tabular*}
	\label{table1}
\end{table}
\subsubsection{Different Numbers of Annotated Samples.}
The influence of different numbers of annotated samples is also investigated (see Table \url{A2} of the appendix). We observe that the more annotated samples, the better the anomaly detection performance, although the difference is not huge and the performance gradually reaches bottleneck. Balancing the anomaly detection performance and the cost of annotating lesions, we still use only one single annotated anomalous image.

\subsubsection{Data Augmentation Operations.}
In this experiment, we fix the randomly resampled lesions and their to-be-pasted positions, and then evaluate the importance of six augmentation operations and their compositions. From the top panel of Table~\ref{table1}, we find that brightness works much better than each of the other five operations. As shown in the bottom panel of Table~\ref{table1}, the composition of five augmentation operations other than color distortion achieves the highest AUC of 0.8126, which is even higher than that from using 10 annotated samples (an AUC of 0.8052). This clearly indicates the importance of data augmentation. We conjecture it is because DR lesions are tightly linked to the color information and color distortion may significantly destroy important lesion-related color information. So we apply a composition of the five augmentation operations, namely Flip, Contrast, Rotation, Resize, and Brightness, in all subsequent experiments.

%\begin{table}[htbp]
%	\caption{Impact of different MixUp coefficients. \textit{0 vs 1} means to classify DR grade 0 images from grade 1 and \textit{0 vs all} means to classify grade 0 images from all other grades in the testing set.}
%	\centering
%	\begin{tabular*}{\hsize}{@{}@{\extracolsep{\fill}}lcccccccccc}	
%		\hline\multirow{2}{0.4in}{MixUp coef. $\lambda$}& \multicolumn{2}{c}{0 vs 1} & \multicolumn{2}{c}{0 vs 2} & \multicolumn{2}{c}{0 vs 3} & \multicolumn{2}{c}{0 vs 4} & \multicolumn{2}{c}{0 vs all} \\
%		\cline{2-11} & AUC & Acc & AUC & Acc& AUC & Acc & AUC & Acc & AUC & Acc\\
%		\hline 1& $0.7174$ & $0.6648$ & $0.8490$ & $0.7921$& $0.9574$& $0.8922$ & $0.9536$ & $0.8769$ & $0.8126$& $0.8262$\\
%		\hline 0.8& $0.7287$ & $0.6721$ & $0.8487$ & $0.7869$& $0.9558$& $0.8847$ & $0.9541$ & $0.8923$ & $0.8150$& $0.8010$\\
%		0.7& $0.7242$ & $0.6710$ & $0.8455$ & $0.7832$& $0.9581$& $0.8596$ & $0.9546$ & $0.8769$ & $0.8177$& $0.8055$\\
%		0.5& $0.7135$ & $0.6304$ & $0.8516$ & $0.7784$& $\mathbf{0.9586}$& $\mathbf{0.9048}$ & $\mathbf{0.9631}$ & $0.8769$ & $0.8096$& $0.8219$\\
%		Random & $\mathbf{0.7348}$ & $\mathbf{0.6778}$ & $\mathbf{0.8528}$ & $\mathbf{0.7932}$& $0.9582$& $0.8947$ & $0.9607$ & $\mathbf{0.9154}$ & $\mathbf{0.8216}$& $\mathbf{0.8254}$\\
%		\hline
%	\end{tabular*}
%	\label{table2}
%\end{table}
\subsubsection{MixUp Coefficients.}
After identifying the optimal data augmentation strategy, the impact of different MixUp coefficients is analyzed. As shown in Table~\ref{table2}, four different MixUp coefficients (three fixed and one random) are tested and the random MixUp coefficient $\lambda \sim U(0.5, 0.8)$ achieves the best performance.

\begin{table}[htbp]
	\caption{The impact of different MixUp coefficients on EyeQ. \textit{0 vs 1} means to classify images of grade 0 from images of grade 1 and \textit{0 vs all} means to classify images of grade 0 from images of all other grades (1-4).}
	\centering
	\begin{tabular*}{\hsize}{@{}@{\extracolsep{\fill}}lccccc}	
		\hline \multirow{2}{0.7in}{MixUp coefficient $\lambda$}&\multicolumn{5}{c}{AUC} \\
		\cline{2-6}& 0 vs 1 &0 vs 2 & 0 vs 3 & 0 vs 4 & 0 vs all \\
		\hline w.o. MixUp& $0.7174$ & $0.8490$ & $0.9574$ & $0.9536$  & $0.8126$\\
		0.8& $0.7287$ & $0.8487$ & $0.9558$ & $0.9541$  & $0.8150$\\
		0.7& $0.7242$ & $0.8455$ & $0.9581$& $0.9546$  & $0.8177$\\
		0.5& $0.7135$ & $0.8516$ & $\mathbf{0.9586}$ & $\mathbf{0.9631}$ & $0.8096$\\
		Random & $\mathbf{0.7348}$& $\mathbf{0.8528}$ & $0.9582$& $0.9607$ & $\mathbf{0.8216}$\\
		\hline
	\end{tabular*}
	\label{table2}
\end{table}

\begin{table}[htbp]
	\caption{Performance comparisons on EyeQ and MosMed.}
	\centering
	\begin{tabular*}{\hsize}{@{}@{\extracolsep{\fill}}lccccccc}
		\hline \multirow{3}*{Methods }& \multirow{3}{0.58in}{$\#$Anomalous samples in training}&\multicolumn{5}{c}{EyeQ}&MosMed\\
		\cline{3-7} \cline{8-8}
		& & \multicolumn{6}{c}{AUC} \\
		\cline{3-7} \cline{8-8}
		&& 0 vs 1 & 0 vs 2 & 0 vs 3 & 0 vs 4 & 0 vs all  & \\
%		\hline && \multicolumn{5}{c}{\emph{ Fully Supervised} }  \\
		\cline{1-7} \cline{8-8}
		\color{gray}{Fully Supervised VGG16} &\color{gray}{1807} &$\color{gray}{0.6257}$ & $\color{gray}{0.8448}$ & $\color{gray}{0.9714}$ & $\color{gray}{0.9846}$ & $\color{gray}{0.7885}$ &$\color{gray}{0.9975}$\\
%		\hline &&\multicolumn{5}{c}{ \emph{Anomaly Detection} } \\
		\cline{1-7} \cline{8-8}
		 f-AnoGAN \cite{f-AnoGAN}&$0$ & $0.5081$ & $0.4915$ & $0.5259$ & $0.5779$ & $0.5148$ &$0.9004$\\
%		MKD \cite{knowledge_distil}&$0$  & $0.5834$ & $0.5493$ & $0.6290$ & $0.7143$ & $0.5576$ \\
		DREAM \cite{dream}&$0$ & $0.5825$ & $0.6655$ & $0.7618$ & $0.7373$ & $0.6252$ &$0.9162$\\
		DevNet  \cite{few-short1}&$10$& $0.5541$ & $0.6530$ & $0.9118$ & $0.9153$ & $0.6301$ &$0.8945$\\
		\cline{1-7} \cline{8-8}
		LesionPaste w.o. MixUp&$1$ & $0.7174$& $0.8490$ & $0.9574$ & $0.9536$ & $0.8126$ &$0.9441$\\
		LesionPaste&$1$  & $\mathbf{0.7348}$ & $\mathbf{0.8528}$ & $\mathbf{0.9582}$ & $\mathbf{0.9607}$ & $\mathbf{0.8216}$&$\mathbf{0.9546}$\\
		\hline
	\end{tabular*}
	\label{table3}
\end{table}	

\subsection{Comparison with State-of-the-art}
In Table~\ref{table3}, we compare our LesionPaste method with state-of-the-art anomaly detection works. As shown in that table, our proposed LesionPaste significantly outperforms all unsupervised learning and semi-supervised learning methods under comparison, and even works better than the fully supervised counterpart in detecting DR of grade 1 with an AUC of 0.7348, grade 2 with an AUC of 0.8528, and all 1-4 grades combined with an AUC of 0.8216. Particularly for detecting DR of grade 1, dramatic improvements of LesionPaste over other methods are observed: an increase of 0.1893 on AUC over the 10-shot anomaly detection method DevNet \cite{few-short1} and an increase of 0.1091 on AUC over the fully supervised method. DR images of grade 1 contain only MA lesions which are extremely tiny in fundus images, and therefore DR of grade 1 is the most challenging anomaly to detect. However, in our LesionPaste, most of the synthesized DR images (80\%) contain only MA lesions, forcing the classification CNN to learn the most difficult samples, so as to improve the performance on detecting DR images of grade 1. In Table~\ref{table3}, LesionPaste also achieves the best result on MosMed. Statistically significant superiority of LesionPaste has been identified from DeLong tests \cite{DeLong_test} at a p-value of $e^{-10}$. Visualization results of the two anomaly detection tasks are shown in Fig. \url{A3}. These results clearly demonstrate the applicability of LesionPaste to different anomaly detection tasks involving different types of diseases, different types of lesions, as well as different types of medical images.

\section{Conclusion}
In this paper, we propose a novel OSAD framework for medical images, the key of which is to synthesize artificial anomalous samples using only one annotated anomalous sample. Different data augmentation and pasting strategies are examined to identify the optimal setting for our proposed LesionPaste. Compared with state-of-the-art anomaly detection methods, either under the unsupervised setting or the semi-supervised setting, LesionPaste shows superior performance on two medical image datasets, especially in the detection of early-stage DR, which even significantly outperforms its fully-supervised counterpart.

%\subsubsection{Acknowledgements} 
%This study was supported by the Shenzhen Basic Research Program (JCYJ20190809120205578); the National Natural Science Foundation of China (62071210); the Shenzhen Basic Research Program (JCYJ202009-
%25153847004); the High-level University Fund (G02236002).


\newpage
\appendix   %仅一个附录时用appendix，否则\appendices
\setcounter{table}{0}   %从0开始编号，显示出来表会A1开始编号
\setcounter{figure}{0}
%定义编号格式，在数字序号前加字符“A"
\renewcommand{\thetable}{A\arabic{table}}
\renewcommand{\thefigure}{A\arabic{figure}}

\begin{figure*}[htbp]
	\centering
	\includegraphics[width=9cm]{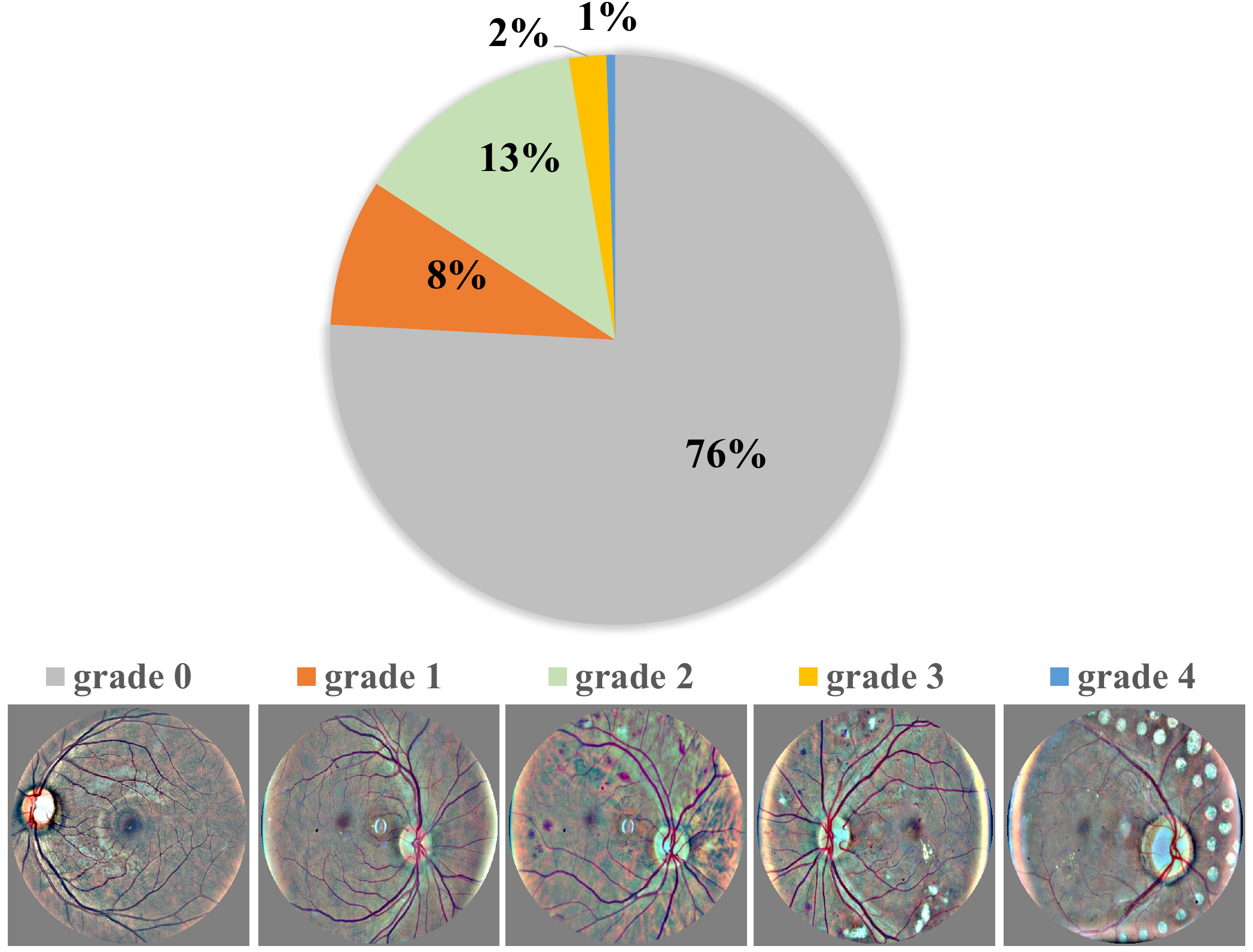}
	%  \vspace{2.0cm}
	\caption{Class distribution and representative preprocessed images of the five categories (grades 0-4) of the training set in EyeQ.}
	\label{figureA1}
\end{figure*}
\begin{figure*}[htbp]
	\centering
	\includegraphics[width=7cm]{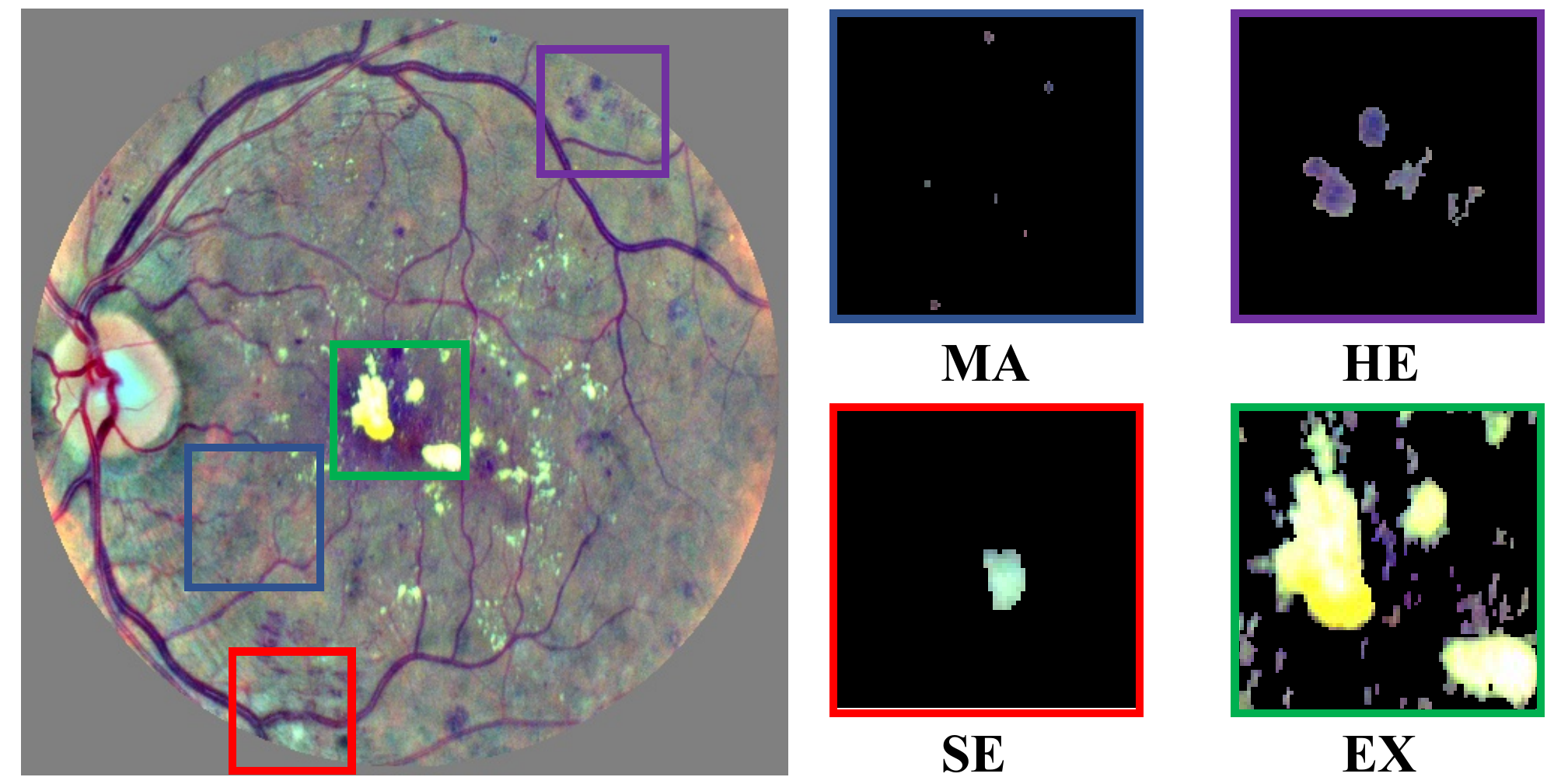}
	%  \vspace{2.0cm}
	\caption{A representative preprocessed fundus image with four types of DR related lesions.}
	\label{figureA2}
\end{figure*}

\begin{table}[htbp]
	\caption{The impact of using different annotated samples. \textit{0 vs 1} means to classify images of grade 0 from images of grade 1 and \textit{0 vs all} means to classify images of grade 0 from images of all other grades (1-4).}
	\vspace{5pt}
	\centering
	\begin{tabular*}{\hsize}{@{}@{\extracolsep{\fill}}lccccc}	
		\hline\multirow{2}{0.7in}{IDRiD image index}& \multicolumn{5}{c}{AUC}\\
		\cline{2-6} & 0 vs 1 & 0 vs 2 & 0 vs 3 & 0 vs 4 & 0 vs all \\
		\hline idrid\_03& $0.6694$ & $0.8293$ & $0.9509$& $0.9562$ & $0.7826$\\
		idrid\_13& $0.6064$ & $0.8181$ & $0.9696$& $0.9640$ & $0.7576$\\
		idrid\_39& $0.6253$ & $0.8033$ & $0.9624$& $0.9505$ & $0.7545$\\
		idrid\_48& $\mathbf{0.6618}$ & $\mathbf{0.8360}$ & $\mathbf{0.9752}$ & $\mathbf{0.9659}$ & $\mathbf{0.7851}$\\
		idrid\_49& $0.6532$ & $0.8325$ & $0.9651$& $0.9631$ & $0.7797$\\
		\hline
	\end{tabular*}
	\label{tableA1}
\end{table}

\begin{table}[htbp]
	\caption{The impact of using different numbers of annotated samples on EyeQ. \textit{0 vs 1} means to classify images of grade 0 from images of grade 1 and \textit{0 vs all} means to classify images of grade 0 from images of all other grades (1-4).}
	\vspace{5pt}
	\centering
	\begin{tabular*}{\hsize}{@{}@{\extracolsep{\fill}}lccccc}	
		\hline\multirow{2}{1.1in}{Number of annotated samples}& \multicolumn{5}{c}{AUC} \\
		\cline{2-6}& 0 vs 1 & 0 vs 2 & 0 vs 3 & 0 vs 4 & 0 vs all \\
		\hline 1&$0.6618$ & $0.8360$ & $0.9752$& $0.9659$ & $0.7851$\\
		2& $0.6784$ & $0.8450$ & $0.9527$ & $0.9662$  & $0.7942$\\
		3& $0.6820$ & $0.8416$ & $0.9627$& $0.9572$ & $0.7992$\\
		5& $0.6836$ & $0.8493$ & $0.9677$ & $\mathbf{0.9733}$  & $0.8011$\\
		10 & $\mathbf{0.6911}$ & $\mathbf{0.8556}$ & $\mathbf{0.9651}$& $0.9699$& $\mathbf{0.8052}$\\
		\hline
	\end{tabular*}
	\label{tableA2}
\end{table}
\begin{figure*}[htbp]
	\centering
	\includegraphics[width=12.5cm]{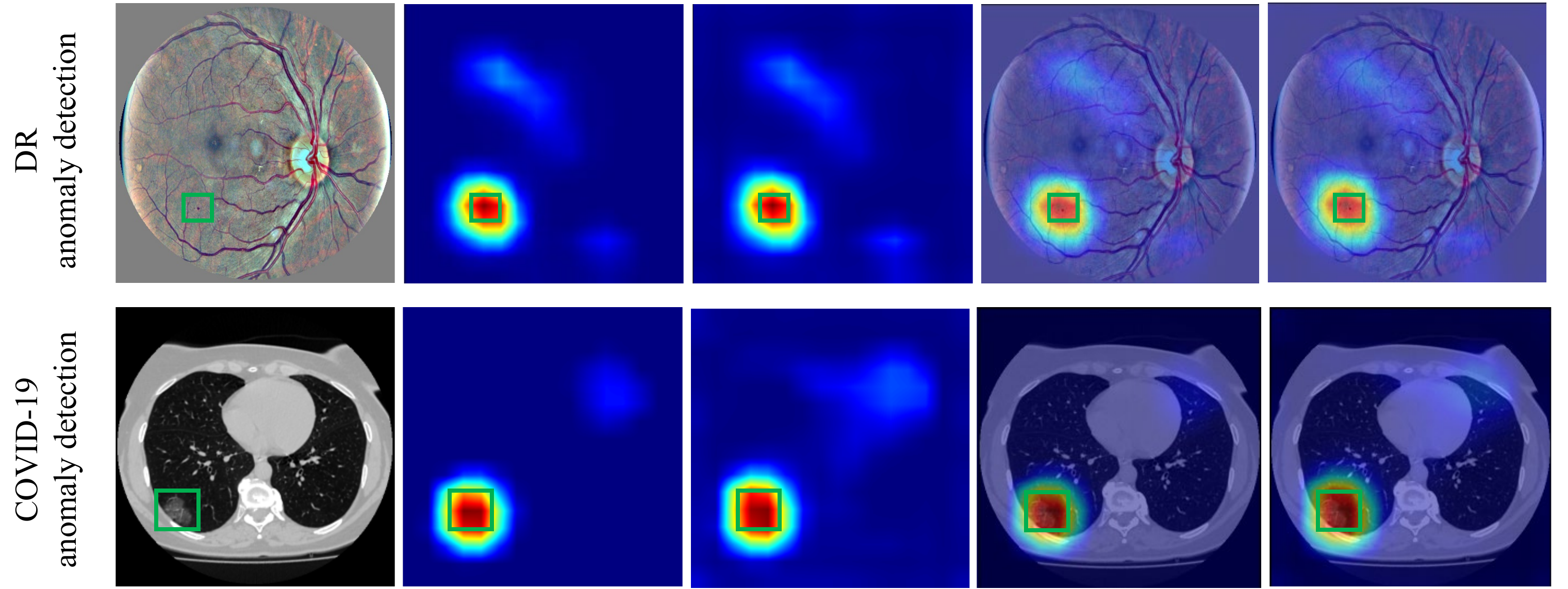}
	%  \vspace{2.0cm}
	\caption{Visualization of representative anomaly detection results. From left to right: abnormal image, GradCAM heatmap, GradCAM++ heatmap, GradCAM heatmap superimposed on the original image, and GradCAM++ heatmap superimposed on the original image. The green bounding boxes highlight lesions/anomalies.}
	\label{figureA3}
\end{figure*}

\end{document}